\documentclass[reprint,
 amsmath,amssymb,
 aps, superscriptaddress,prd
]{revtex4-2}
\usepackage{graphicx}
\usepackage{dcolumn}
\usepackage{bm}
\usepackage{relsize}
\usepackage{fontenc}
\usepackage{amssymb}
\usepackage{tabularx}
\usepackage{tikz}
\usepackage{pgfplots}
\usepackage{mathtools}
\usepackage{siunitx}
\usepackage{physics}
\usepackage[compat=1.1.0]{tikz-feynman}
\usepackage[T1]{fontenc}
\usepackage{multirow}
\usepackage{subfigure}
\usepackage{booktabs}
\usepackage{soul}
\usepackage{mathrsfs}

\def\({\left(} 
\def\){\right)}

\newcommand{\smallw}{{\scriptscriptstyle W}}

\newcommand{\sw}{s_{\smallw}} 
\newcommand{\cw}{c_{\smallw}}

\newcommand{\Amp}{\mathcal{M}}

\definecolor{mypurple}{rgb}{0.458, 0.078, 0.96} \definecolor{myblue}{rgb}{0.32, 0.70, 0.90} \definecolor{mygreen}{rgb}{0.63, 0.988, 0.73} \definecolor{myyellow}{rgb}{0.956, 0.713, 0.43} %
\usepackage{hyperref}
\begin{document}

\title{\boldmath New physics contamination to precision luminosity measurements \\ at future $e^+e^-$ colliders}

\author{Mauro Chiesa}
\email{mauro.chiesa@pv.infn.it}
\affiliation{INFN, Sezione di Pavia, Via A. Bassi 6, 27100 Pavia, Italy}
\author{Clara L. Del Pio}
\email{cdelpio@bnl.gov}
\affiliation{Department of Physics, Brookhaven National Laboratory,
Upton, New York 11973, USA}
\author{Guido Montagna}
\email{guido.montagna@unipv.it}
\affiliation{Dipartimento di Fisica "Alessandro Volta", Universit\`a di Pavia, Via A. Bassi 6, 27100 Pavia, Italy}
\affiliation{INFN, Sezione di Pavia, Via A. Bassi 6, 27100 Pavia, Italy}
\author{Oreste Nicrosini}
\email{oreste.nicrosini@pv.infn.it}
\affiliation{INFN, Sezione di Pavia, 
Via A. Bassi 6, 27100 Pavia, Italy}
\author{Fulvio Piccinini}
\email{fulvio.piccinini@pv.infn.it}
\affiliation{INFN, Sezione di Pavia, Via A. Bassi 6, 27100 Pavia, Italy}
\author{Francesco P. Ucci}
\email{francesco.ucci@pv.infn.it}
\affiliation{Dipartimento di Fisica "Alessandro Volta", Universit\`a di Pavia, Via A. Bassi 6, 27100 Pavia, Italy}
\affiliation{INFN, Sezione di Pavia, Via A. Bassi 6, 27100 Pavia, Italy}

\begin{abstract}
    Several key observables of the high-precision physics program at future lepton colliders will critically depend on the knowledge of the absolute machine luminosity. 
The determination of the luminosity relies on the precise knowledge of some reference process, which is, in principle, not affected by unknown physics, so that its cross section can be computed within a well-established theory, like the Standard Model. Quantifying the uncertainties induced by possible New Physics effects on such processes is, therefore, crucial. We present an investigation of light and heavy 
new physics contributions to the small-angle Bhabha process at future $e^+e^-$ colliders and we discuss possible strategies to remove the contamination due to heavy degrees of freedom by relying on 
observables that are independent of the absolute 
luminosity.

\end{abstract}

\maketitle

\section{Introduction}
\label{sec:intro}
The absolute luminosity measurement is extremely important at $e^+ e^-$ colliders, from low to 
high energy, because it is a source of 
systematics in the measurement of the cross section of signal processes. For instance, at the few-GeV scale
of flavor factories, it is crucial for the determination of the pion form factor through the measurement of the hadronic cross section in $e^+ e^-$ collisions~\cite{WorkingGrouponRadiativeCorrections:2010bjp}, while at the energies of 
$Z$ factories, like LEP in the past and at Higgs, top and electroweak factories in the future~\cite{Altmann:2025feg}, it is relevant for the determination of some key Standard Model (SM) parameters~\cite{ALEPH:2005ab,Voutsinas:2019hwu,Janot:2019oyi}.
An example of the importance of a precise knowledge of the absolute luminosity is given by the recent study of the effects of beam-beam interactions at LEP~\cite{Voutsinas:2019hwu}, where an underestimate of the LEP luminosity of about 0.1\% was discovered, removing a long-standing tension on the number of light neutrino species from LEP~\cite{Janot:2019oyi}.

At future $e^+ e^-$ high energy colliders, the luminosity calibration will also play a crucial role 
in improving the precision of electroweak (EW) measurements. For instance, an uncertainty of $10^{-4}$ on the luminosity at $WW$ threshold would allow the precise knowledge of the $W$ mass and width with a $0.3$ and $1.2~{\rm{MeV}}$ statistical precision respectively, through the measurement of the $W$-pair cross section line shape~\cite{azzi2017physicsprecision,Azzurri_2021}. Moreover, an unprecedented precision on the effective $HZZ$ coupling and the total Higgs boson width through the measurement of the total cross section of the process $e^+ e^- \to H Z$ at energy scales above the $HZ$ threshold could be achieved if the systematic error due to luminosity is kept small, in order to fully exploit the achievable statistics. 

The time-integrated luminosity $L$ is related to the cross section $\sigma_0$ of some reference process through the relation 
\begin{equation}
L = \int {\cal L} \, \dd t = \frac{N_0}{ \epsilon  \sigma_0} \, ,
\label{eq:luminositydef}
\end{equation} 
where ${\cal L}$ is the instantaneous luminosity, $N_0$ is the number of observed events of such a process and $\epsilon$ is the experimental selection efficiency. On 
the one hand, the choice of the reference processes is motivated by clean experimental signatures and large cross sections, in order to minimise the experimental systematics. On the other hand, a further important requirement is the feasibility of a very precise theoretical description of the differential cross sections of the reference process, since the precision in the cross section calculation enters directly as a source 
of systematics.
\begin{figure}[t]
\begin{equation*}
    \begin{gathered}
\begin{tikzpicture}
  \begin{feynman}[small]
    \vertex (a);
    \vertex[right=1.25cm of a] (b);
    \vertex[above left=1cm and 0.375cm of a] (c) {$e^+$};
    \vertex[below left=1cm and 0.375cm of a] (d) {$e^-$};
    \vertex[above right=1cm and 0.375cm of b] (e) {$e^+$};
    \vertex[below right=1cm and 0.375cm of b] (f) {$e^-$};
    \diagram* {
      (a) -- [photon,edge label=$\gamma\,\text{,}\,Z$] (b),
      (d) -- [fermion] (a),
      (a) -- [fermion] (c),
      (e) -- [fermion] (b) --[fermion] (f),
    };
  \end{feynman}
\end{tikzpicture}
    \end{gathered}
    \quad\qquad
    \begin{gathered}
     \begin{tikzpicture}
    \vspace{2mm}
  \begin{feynman}[small]
    \vertex (a);
    \vertex[below=1.25cm of a] (b);
    \vertex[above left=0.375cm and 1cm of a] (c) {$e^+$};
    \vertex[below left=0.375cm and 1cm of b] (d) {$e^-$};
    \vertex[above right=0.375cm and 1cm of a] (e) {$e^+$};
    \vertex[below right=0.375cm and 1cm of b] (f) {$e^-$};
    \diagram* {
      (a) -- [photon,edge label=$\gamma\,\text{,}\,Z$] (b),
      (d) -- [fermion] (b),
      (a) -- [fermion] (c),
      (e) -- [fermion] (a),
      (b) --[fermion] (f),
    };
  \end{feynman}
\end{tikzpicture}
    \end{gathered}
\end{equation*}
\caption{Tree level $s$- and $t$-channel diagrams for the Bhabha scattering.}
\label{fig:SMdiags}
\end{figure}
At LEP, the above conditions were satisfied by small angle Bhabha scattering (SABS), where the process is largely dominated by the QED photon
exchange and the imperfect knowledge of the weak interactions has not been a matter of concern. This can be understood by looking at the angular dependence of the SABS cross section differential in the electron scattering angle that is given by
\begin{equation}
\dv{\sigma_0}{\theta} = \frac{32\pi \alpha^2}{s \theta^3} \left( 1 - \frac{\theta^2}{2} + \frac{9}{40} \theta^4 + \delta_{\text{weak}} \right),
\end{equation}
where the $1/\theta^3$ behavior is due to the photon $t$-channel diagram shown in the right panel in Fig.~\ref{fig:SMdiags} and $s$ is the center of mass (c.m.) energy. In the LEP luminometer acceptance, the contribution due to the $Z$ exchange was at most $\delta_\text{weak}\sim0.3\%$~\cite{Arbuzov:283383}, with its radiative corrections being at the level of $10^{-4}$. At flavor factories, the main reference process is the large-angle 
Bhabha scattering (LABS)~\cite{WorkingGrouponRadiativeCorrections:2010bjp}
and the processes $e^+ e^- \to \mu^+ \mu^-$ and $e^+ e^- \to \gamma \gamma$ are also used for 
normalization and cross-checks.

The highest relative experimental precision in the absolute luminosity calibration was achieved by the OPAL Collaboration~\cite{OPAL:1999clt} at LEP, with the value of $3.4 \times 10^{-4}$. Such a high level of precision was matched by theoretical predictions via the inclusion of exact next-to-leading-order (NLO) calculations and the resummation of higher-order photon emissions~\cite{Montagna:1998sp}. 
At future $e^+ e^-$ high-energy colliders, the precision requirements for the luminosity measurements will be even more demanding, being of the order of at least $10^{-4}$, or possibly better, at the $Z$ resonance and at the $WW$ threshold region, and of the order of $10^{-3}$ at c.m. energies larger than about $240$~GeV~\cite{deBlas:2024bmz}. 
On the theoretical side, studies on the needs for the SM calculations and Monte Carlo event generators to reach such unprecedented  precisions have appeared recently in the literature~\cite{Jadach:2018jjo,Proceedings:2019vxr,Ward:2019ooj,Ward:2020yif,Jadach:2021ayv,Skrzypek:2024gku,Ward:2024frh}. 
While this outlines a clear program for improving SM calculations, for example including also NLO weak corrections, the impact of new physics (NP) on luminosity measurements remains less explored. An unknown degree of freedom (d.o.f.) could contribute to the benchmark process cross section, potentially biasing absolute cross section determinations or at least increasing the associated theoretical uncertainty. 

In this paper, we investigate whether NP could contaminate the absolute luminosity measurements at various future high-energy $e^+ e^-$ 
colliders proposals through SABS, 
using the present and projected knowledge of the bounds on beyond the Standard Model (BSM) effects.  
A complementary investigation in this direction was presented in 
\cite{Maestre:2022cvs} for the process $e^+ e^- \to \gamma \gamma$, which is presently under consideration as an alternative luminosity process~\cite{CarloniCalame:2019dom,Dam:2021sdj,deBlas:2024bmz}, highlighting the interplay between higher-dimension NP effects and FCC-ee luminosity measurements. Additionally, we explore the effectiveness of the asymmetry measurements of LABS as a tool to remove the uncertainties due to possible NP contributions.
%effectively drawing a possible running strategy \textbf{non so se mettere questo commento sulla running strategy, che può risultare molto "teorica"}.

\section{New physics impact on SABS}
\label{sec:th-intro}
The NP contribution to the SABS cross section can be due to heavy new d.o.f. or to light mediators with feeble couplings to the leptons or photons which have escaped  detection until now. For the former case, the natural framework for a model-independent analysis is the Standard Model effective field theory (SMEFT), where all possible higher-dimensional operators are added to the SM Lagrangian. For the latter case of light NP contributions, a model-dependent analysis is necessary. We will, therefore, study the effects due to the exchange of light d.o.f. with spin $0$ or $1$ and with different parity. 

We consider in this analysis various future $e^+ e^-$ collider options, in particular FCC~\cite{FCC:2018byv,FCC:2018evy,FCC:2025lpp}, CEPC~\cite{CEPCStudyGroup:2018rmc,CEPCStudyGroup:2018ghi}, ILC~\cite{Behnke:2013xla,ILC:2013jhg,Adolphsen:2013jya,Adolphsen:2013kya,Behnke:2013lya,Jelisav_i__2013}, and CLIC~\cite{Linssen:2012hp,CLIC:2016zwp,CLICdp:2018cto,cern_edms_1644637483}. The typical angular acceptances of the luminometers~\footnote{We remark that the fiducial volume for SABS events is smaller than the geometrical acceptance of the luminometer.} and c.m. energies of the colliders are specified in the second and third columns in Table~\ref{tab:SMEFT_numbers}, respectively. 
For the EW sector, we adopt the $\{\alpha(M_Z),G_\mu, M_Z\}$ input parameter scheme with the following numerical values
\begin{align*}
    \alpha(M_Z)&=1/127.95\\
    G_\mu&=\SI{1.16638d-5}{\giga\electronvolt\tothe{-2}}\\
    M_Z&=\SI{91.1876}{\giga\electronvolt}\, .
\end{align*}
The relevant amplitudes have been generated with \textsc{FeynArts}~\cite{Hahn:2000kx} and \textsc{FeynCalc}~\cite{Mertig:1990an,Shtabovenko:2016sxi}. For the SMEFT, we have used the UFO~\cite{Darm__2023} model from \textsc{SmeftFR}~\cite{Dedes:2017zog,Dedes:2019uzs,Dedes:2023zws} and checked against \textsc{SMEFTSim}~\cite{Brivio:2017btx,Brivio:2020onw} within \textsc{MG5\_aMC@NLO}~\cite{Alwall:2014hca}, while the UFO used for light NP has been obtained by implementing the Lagrangians in \textsc{FeynRules}~\cite{Alloul:2013bka} and used as model files in ~\textsc{FeynArts}.

 Numerical results have been obtained with an updated version~\footnote{The running code and all the results are available on \href{https://github.com/francescoucci/luminosity/}{GitHub}} of the \textsc{BabaYaga@NLO} Monte Carlo generator~\cite{CarloniCalame:2000pz,CarloniCalame:2001ny,Balossini:2006wc,Balossini:2008xr,Budassi:2024whw}, able to simulate the Bhabha scattering with NLO QED corrections matched to a parton shower for $\gamma$ and $Z$ exchange. The latest version includes NP effects for light and heavy mediators as described in the following.
\subsection{Light new physics scenarios}
If the mass of NP is below the EW scale, i.e. $M_{\rm{NP}}\lesssim \Lambda_{\rm{EW}}$,  an EFT approach is not suitable and we need to rely on specific renormalizable models. Inspired by~\cite{Masiero:2020vxk}, we analyze the maximal contamination of couplings of SM particles to (pseudo)scalar, (axial)vector 
NP d.o.f. that are not excluded by the current experimental bounds.
As shown in the following, the numerical results obtained with \textsc{BabaYaga@NLO}, for processes interfering with the SM, can be understood by considering the analytical estimate of the leading light NP (LNP) deviations given by
\begin{equation}\label{Eq:LNPdev}
    \delta_{\rm{LNP}}\simeq \frac{2 \Re \left(\mathcal{M}_\gamma(t)^\dagger \mathcal{M}_{\rm{LNP}}\right)}{|\mathcal{M}_\gamma(t)|^2} \, ,
\end{equation}
where the subscript $\gamma$ indicates the photon exchange, which dominates the SM SABS amplitude in the $t$ channel, whereas $\mathcal{M}_\text{LNP}$ is the amplitude for the considered light NP models, whose relevant Feynman diagrams are represented in Fig.~\ref{fig:LNPdiags}.
\begin{figure}[h]
\begin{equation*}
    \begin{gathered}
\begin{tikzpicture}
  \begin{feynman}[small]
    \vertex[style=dot,fill=myblue, draw=myblue] (a) {};
    % \vertex[below=1.25cm of a ,style=dot, label={[yshiftbelow:$g_{aee}$}] (b) {};
\vertex[right=1.25cm of a ,style=dot, draw=myblue, fill=myblue] (b) {};
    
    \vertex[above left=1cm and 0.375cm of a] (c) ;
    \vertex[above right=1cm and  0.375cm  of b] (d) ;
    \vertex[below left=1cm and  0.375cm  of a] (e);
    \vertex[below right=1cm and  0.375cm  of b] (f);
    \diagram* {
      (a) -- [scalar,edge label=$a$] (b),
      (d) -- [fermion] (b),
      (a) -- [fermion] (c),
      (e) -- [fermion] (a),
      (b) --[fermion] (f),
    };
  \end{feynman}
\end{tikzpicture}
    \end{gathered} 
    \quad\qquad
    \begin{gathered}
     \begin{tikzpicture}
  \begin{feynman}[small]
    \vertex (a);
    \vertex[below=1.5cm of a] (b);
    \vertex[above left=0.25cm and 1cm of a] (c);
    \vertex[below left=0.25cm and 1cm of b] (d) ;
    \vertex[above right=0.25cm and 1cm of a] (e);
    \vertex[below right=0.25cm and 1cm of b] (f);
    \vertex[below right=0.75 cm and 0.3 cm of a, style= dot,fill=myyellow, draw=myyellow] (g) {};
    \vertex[right=1cm of g] (h) {$a$};
    \diagram* {
      (a) -- [photon] (g),
      (g) -- [photon] (b),
      (d) -- [fermion] (b),
      (a) -- [fermion] (c),
      (e) -- [fermion] (a),
      (b) --[fermion] (f),
      (g) --[scalar] (h),
    };
  \end{feynman}
\end{tikzpicture}
    \end{gathered}
    \qquad   \begin{gathered}
\begin{tikzpicture}
  \begin{feynman}[small]
    \vertex[style=dot, fill=mypurple,draw=mypurple] (a) {};
    \vertex[below=1.25cm of a ,style=dot,fill=mypurple,draw=mypurple] (b) {};
    \vertex[above left=0.375cm and 1cm of a] (c) ;
    \vertex[below left=0.375cm and 1cm of b] (d) ;
    \vertex[above right=0.375cm and 1cm of a] (e);
    \vertex[below right=0.375cm and 1cm of b] (f);
    \diagram* {
      (a) -- [boson,edge label=$V\text{,}\,X_{17}$] (b),
      (d) -- [fermion] (b),
      (a) -- [fermion] (c),
      (e) -- [fermion] (a),
      (b) --[fermion] (f),
    };
  \end{feynman}
\end{tikzpicture}
    \end{gathered}
\end{equation*}
\caption{Diagrams for LNP contributions to SABS. To the left and right, the $t$-channel diagram mediated either by a scalar or a dark vector, respectively. The central diagram represents the photon-photon fusion producing a light ALP in the final state. Dotted vertices correspond to LNP couplings, indicated by colors.}
\label{fig:LNPdiags}
\end{figure}

\subsubsection{(Pseudo)scalar axionlike particles}
The interaction of a (pseudo)scalar axionlike particle (ALP) $a$ of mass $m_a$ with both the photon and the electron can be parametrized with the parity-violating Lagrangian 
\begin{equation}
\begin{aligned}
    \mathscr{L}_\text{ALP} =&\frac{1}{2}\partial_\mu a \, \partial^\mu a -\frac{1}{2} m_a^2 a^2\\
    &+\frac{1}{4}g_{a\gamma\gamma}( F_{\mu\nu}\Tilde{F}^{\mu\nu})\, a+  g_{aee} (\bar{e}\,i\gamma_5 \, e)\, a\, ,
\end{aligned}
\label{eq:three}
\end{equation}
where $F_{\mu\nu}$ is the electromagnetic field tensor, $\Tilde{F}^{\mu\nu}$ the dual 
field tensor and $e$ the electron field. In Eq.~\ref{eq:three},
the scalar parity-conserving case is obtained with the substitutions $\Tilde{F}^{\mu\nu}\to{F}^{\mu\nu}$, $i\gamma_5\to \mathbb{I}$. In the limit $m_a\ll \sqrt{s}$, the leading contribution in the axion-to-electron coupling $g_{aee}$ is given by the interference of the $t$-channel photon exchange with the $s$-channel diagram mediated by the ALP, as $\mathcal{M}^\dagger_\gamma(t)\mathcal{M}_a(t)$ vanishes in the limit $m_e \simeq 0$, yielding a relative differential deviation given by
\begin{equation}
\delta^{aee}_\text{ALP}\simeq \frac{g_{aee}^2}{4 \pi  \alpha}\frac{s^2 t}{ \left(s-m_a^2\right) \left(s^2+u^2\right)} 
 \simeq -\frac{g_{aee}^2}{8 \pi  \alpha}(1-\cos\theta) \, ,
\end{equation}
which is highly suppressed at small angle.

In order to evaluate the maximal effect, we consider the bounds put by the NA64 beam-dump experiment via the detection of missing energy in the hard bremsstrahlung process $e^- Z \to e^- Z X \, ; \, X \to \text{invisible}$~\cite{NA64:2021xzo} where $X$ can have different intrinsic spin. In the unconstrained mass-coupling plane, we take the point with the largest value of the coupling of electrons to the axion, given by $\left(g_{aee},m_a\right)\simeq(3\times 10^{-3},\SI{1}{\giga\electronvolt})$. At small angle, this yields $\delta^{aee}_\text{ALP}<{10^{-7}}$, well below any future collider precision goal.
A light ALP can be also produced via photon fusion in the process $e^+e^- \to \gamma^* \gamma^* e^+ e^- \to e^+e^- a$, shown in the middle panel in Fig.~\ref{fig:LNPdiags}, that can give the same experimental signature of 
SABS~\cite{Acanfora:2023gzr}. For this process, the results have been obtained with a modified version of the code developed for simulating $\pi^0 \pi^0$~\cite{Nguyen:2006sr} and 
$\eta$~\cite{KLOE-2:2012lws} 
production in $e^+ e^-$ collisions and $\pi^0$ production in $\mu e$ scattering~\cite{Budassi:2022kqs}. For the largest admitted value of the photon-axion coupling $g_{a\gamma\gamma}=\SI{2e-4}{\giga\electronvolt\tothe{-1}}$ obtained via beam-dump experiments and invisible decay searches, as reported in Ref.~\cite{Acanfora:2023gzr} and references therein, this contribution at the $Z$ peak in a small angular acceptance is suppressed, giving $\delta^{a\gamma\gamma}_\text{ALP}\sim \order{10^{-6}}$. At higher energy, the bulk of the axion emission cross section is unchanged~\cite{Brodsky:1971ud}, while the SM cross section falls as $1/s$, yielding  $\delta^{a\gamma\gamma}_\text{ALP}(\SI{1}{\tera\electronvolt})\sim \order{10^{-5}}$. 
This feature can be understood since the total cross section for axion emission grows logarithmically with the c.m. energy~\cite{Brodsky:1971ud} but the angular distribution of the electron and positron around the beam axis becomes more sharply peaked. The fixed angular acceptance effectively removes the increase of the cross  section at high energies. 
\subsubsection{Dark vectors}
In a light NP scenario, Bhabha scattering could also be mediated by a dark vector boson ${ V_{\mu}}$ associated with a new $U(1)'$ gauge symmetry, usually referred to as dark photon~\cite{Bauer:2018onh} or light $Z'$ models~\cite{Nath:2021uqb}. The kinetic mixing of such boson~\cite{Holdom:1985ag,Boehm:2003hm,Pospelov:2007mp} with the SM (un)broken  $U(1)_{(\rm{em})Y}$ gauge field would result in a coupling with the electromagnetic current after the EW symmetry breaking. The Lagrangian we use to estimate such effects is given by 
\begin{equation}\label{eq:LagVec}
\begin{aligned}
    \mathscr{L}_\text{Dark} =&-\frac{1}{4} V^{\mu\nu}V_{\mu\nu} +\frac{1}{2}M_V^2 V_\mu V^\mu\\
    &+g_V' \left(\bar e \, \gamma^\mu \, e\right)V_\mu + g_A' \left(\bar e \, \gamma^\mu\gamma_5 \, e\right)V_\mu \, ,
    \end{aligned}
\end{equation}
where $g'_{V,A}$ are the vector and axial couplings of the dark spin-1 mediator of mass $M_V$ with electrons
and $V_{\mu\nu}$ is the dark field tensor. The bulk of the deviation with respect to the SABS in the SM due to the $V_\mu$ mediator is given by the interference of the QED amplitude with the dark vector $t$-channel exchange, shown in the rightmost panel of Fig.~\ref{fig:LNPdiags}, i.e.
\begin{equation}
\delta_{\rm{Dark}}\simeq\frac{t \left[{g_V'}^{2} \left(s^2+u^2\right)-{g_A'}^{2} \left(s^2-u^2\right)\right]}{2 \pi  \alpha \left(t-M_V^2\right) \left(s^2+u^2\right)}\,.
\end{equation}
At small angles, one has $u\simeq -s$, resulting in a strongly suppressed axial-vector coupling. Bounds on $g_V'$ are taken from the BABAR search for dark photons~\cite{BaBar:2017tiz} via the process $e^+ e^- \to \gamma V$; $V \to $~invisible, which are found to be more constraining than~\cite{NA64:2021xzo} in the region of $M_V\simeq \SI{1}{\giga\electronvolt}$. Therefore, the maximal contribution of the dark vector current is given for $(g_{V}',M_V)\simeq(3\times10^{-4},\SI{1}{\giga\electronvolt})$, resulting in $\delta_{\rm{Dark}}\sim \order{10^{-6}}$, 
 which is 2 orders of magnitude below the foreseen 
luminosity precision of future colliders.

\subsubsection{The hypothetical $X_{17}$ particle}
The anomalous resonance observed by the ATOMKI experiment~\cite{krasznahorkay2015observation} in ${}^8 \text{Be}^*\to{}^8\text{Be}\, e^+e^-$ 
nuclear transitions at $m_{e^+e^-}\simeq17~\rm{MeV}$ has been interpreted as a new vector gauge boson~\cite{feng2016protophobic}. Subsequent results with ${}^4\text{He}$~\cite{krasznahorkay2021new} and ${}^{12}\text{C}$~\cite{krasznahorkay2022new} atoms confirmed such findings, while the MEG II experiment~\cite{MEGII:2024urz} has not found evidence in its data, yet still being compatible with the existence of the $X_{17}$ boson. Recently, the PADME experiment observed an excess $e^+e^-$ data~\cite{bossi2025searchnew17mev}, by searching the resonant production and decay $X_{17}\to e^+e^-$ via a fixed target experiment, being compatible with the ATOMKI average. The experimental findings have been reviewed in detail in Ref.~\cite{fern2025combined} while their phenomenological implications have been studied in~\cite{Barducci_2025,DiLuzio:2025ojt}, but no consensus has been reached in the literature on the parity of the $X_{17}$ coupling to electrons. Therefore, we parametrize the effect of the possible $X_{17}$ mediation in the SABS with the (axial)vector Lagrangian of Eq.~\eqref{eq:LagVec}. We consider the most recent constraints from the PADME experiment that do not exclude the values $(g_{Xee},m_X)=(5.6\times10^{-4},16.90~\text{MeV})$. We find that the deviations due to the $X_{17}$ in the SABS for the FCC-ee setup at $\sqrt{s}=M_Z$ are of the order of $\delta_{X_{17}}\simeq 6\times 10^{-6}$, while at higher energies its effect is below the luminosity precision target.
More data are foreseen the in next years that will shed light on the hypothetical $X_{17}$ particle allowing for a precise estimation of the uncertainties to the SABS associated with it. 

We can conclude that no contribution from light NP is expected to alter the SM picture of the SABS cross section at $10^{-4}$ level, allowing us to neglect such effects in the rest of this work. 
\subsection{Heavy new physics scenarios}
Under the assumption that the NP scale lies far above the electroweak scale, {i.e.} $\Lambda_\text{NP} \gtrsim \order{\mathrm{TeV}}$, one can study deviations from the SM using the SMEFT framework at dimension six (see Ref.~\cite{Brivio:2017vri} for a review), a model-independent way to capture the leading BSM effects. The effective Lagrangian is expanded about the SM as follows 
\begin{equation}
    \mathscr{L}_\text{SMEFT} = \mathscr{L}_\text{SM} + \sum_i \frac{C_i \hat O_i^{(6)}}{\Lambda^2_\text{NP}} + \order{\frac{1}{\Lambda^4_\text{NP}}}\, ,
\end{equation}
where the operators $\hat O_i^{(6)}$ are built from the same d.o.f. of the SM as gauge-invariant combinations under $\mathrm{SU}(3)_C \times \mathrm{SU}(2)_L\times \text{U}(1)_Y$ and $C_i$ are the associated Wilson coefficients (WCs). In the following, we do not make any flavor symmetry assumption, therefore considering a completely general flavor scenario, for which the coefficients $[C_{i}]_J$ come with flavor indices $J$. Since we are dealing only with the first generation of leptons, we suppress such indices unless needed for clarity. Nevertheless, this flavor assumption is crucial in global fits as it reduces the degeneration between coefficients entering the predictions for different processes. 

%In principle, coefficients $C_i^{\{f\}}$ come with flavor indices $\{f\}$ which we suppress since we deal only with electrons; nevertheless this assumption is crucial in global fits as it reduces the degeneration between coefficients entering different processes. 
\begin{figure*}[t]
\begin{equation*}
    \begin{gathered}
    \vspace{-0.5cm}
\begin{tikzpicture}
  \begin{feynman}[small]
            \vertex (a);
            \vertex[below=1.25cm of a ,style=square dot, fill= mypurple, draw=mypurple, minimum size=4pt, label={[yshift=-6pt]below:$\Delta g_{L/R}^{Ze}$}] (b) {};
    \vertex[above left=0.375cm and 1cm of a] (c) ;
    \vertex[below left=0.375cm and 1cm of b] (d) ;
    \vertex[above right=0.375cm and 1cm of a] (e);
    \vertex[below right=0.375cm and 1cm of b] (f);
    \diagram* {
      (a) -- [photon,edge label=$Z$] (b),
      (d) -- [fermion] (b),
      (a) -- [fermion] (c),
      (e) -- [fermion] (a),
      (b) --[fermion] (f),
    };
  \end{feynman}
\end{tikzpicture}
    \end{gathered} 
    \quad\qquad
   \begin{gathered}
\begin{tikzpicture}
  \begin{feynman}[small]
    \vertex[style=square dot, minimum size=4pt,fill=myblue, draw=myblue, label={[yshift=-7pt]below:$C_{i}$}] (a) {};
    \vertex[above left=1cm and 1cm of a] (c) ;
    \vertex[below left=1cm and 1cm of a] (d) ;
    \vertex[above right=1cm and 1cm of a] (e);
    \vertex[below right=1cm and 1cm of a] (f);
    \diagram* {
      (d) -- [fermion] (a),
      (a) -- [fermion] (c),
      (e) -- [fermion] (a),
      (a) --[fermion] (f),
    };
  \end{feynman}
\end{tikzpicture}
    \end{gathered}\qquad \qquad\qquad \begin{gathered}
\begin{tikzpicture}
  \begin{feynman}[small]
    \vertex[style=square dot, minimum size=4pt,fill=myblue, draw=myblue, label={[yshift=-7pt]below:$C_{i}$}] (a) {};
    \vertex[above left=1cm and 1cm of a] (c) ;
    \vertex[below left=1cm and 1cm of a] (d) ;
    \vertex[above right=1cm and 1cm of a] (e);
    \vertex[below right=1cm and 1cm of a] (f);
    \vertex[above left=0.75cm and 0.75cm of a] (g);
    \vertex[above right=0.75cm and 0.75cm of a] (h);
    \diagram* {
      (d) -- [fermion] (a),
      (a) -- [fermion] (c),
      (e) -- [fermion] (a),
      (a) --[fermion] (f),
      (g) --[photon,looseness=0.5] (h),
    };
  \end{feynman}
\end{tikzpicture}
    \end{gathered}
    \qquad
    \begin{gathered}
    \vspace{-10pt}
    \begin{tikzpicture}
  \begin{feynman}[small]
            \vertex (a);
            \vertex[below=1.25cm of a ,style=square dot, fill= myyellow, draw=myyellow,label={[yshift=-3pt]below:$C_{j}$}, minimum size=4pt] (b) {};
    \vertex[above left=0.375cm and 1cm of a] (c) ;
    \vertex[below left=0.375cm and 1cm of b] (d) ;
    \vertex[above right=0.375cm and 1cm of a] (e);
    \vertex[below right=0.375cm and 1cm of b] (f);
    \vertex[below=0.75 cm of a](g);
    \diagram* {
      (a) -- [photon,edge label=$\gamma\text{,}Z$] (g),
      (d) -- [fermion] (b),
      (a) -- [fermion] (c),
      (e) -- [fermion] (a),
      (b) --[fermion] (f),
      (g) --[fermion,half left] (b) --[fermion, half left] (g),
    };
  \end{feynman}
\end{tikzpicture}
\end{gathered}
\end{equation*}
\caption{Diagrams with insertion of SMEFT operators at dimension six, represented by square vertices. From the left to the right: the first diagram represents a shift of the $Zee$ couplings in one of the two vertices. The second diagram is a four-fermion contact interaction. The latter two diagrams are NLO contribution: the first accounts for a virtual SM boson exchange on top of a four-fermion operator, and the rightmost represents a dimension-six NLO insertion. These diagrams are in interference with the SM ones in Fig.~\ref{fig:SMdiags}.}
\label{fig:HNPdiags}
\end{figure*}
The modified EW Lagrangian in the $\{\alpha,G_\mu,M_Z\}$ scheme reads as follows~\cite{Falkowski:2017pss} 
\begin{equation}
\begin{alignedat}{2}
\mathscr{L}^{\text{EW}}_{\text{SMEFT}} = -\sqrt{4\pi\alpha}& \,(&&
\bar{e} \gamma^\mu e) A_\mu \\
+ \frac{\sqrt{4\pi\alpha}}{\sw \cw}&\Biggl[&&\bar{e}_L\gamma^\mu \left(\hat{g}_L+
\frac{\Delta g_L^{Ze}}{\Lambda^2_\text{NP}}\right) e_L\\
&+&&\bar{e}_R \gamma^\mu \left(\hat{g}_R+ \frac{\Delta g_R^{Ze}}{\Lambda^2_\text{NP}}\right) e_R \Biggr] Z_\mu\, ,
\end{alignedat}
\label{eq:eight}
\end{equation}
where $A_\mu$ and $Z_\mu$ are the photon and 
$Z$-boson field, respectively, and $e_{L,R}$ are the left- and right-handed electron fields. We neglect the  operators $\hat{O}_{eW}^{(6)}=(\bar{e}_L\sigma^{\mu\nu}e_R) \sigma^I\phi\, W_{\mu\nu}^I$ and $\hat{O}_{eB}^{(6)}=(\bar{e}_L\sigma^{\mu\nu}e_R) \phi\, B_{\mu\nu}$, that generate dipole interactions after the Higgs acquires a vacuum expectation value $v$. The corresponding WCs $C_{eB,eW}$ are tightly constrained by lepton dipole moments~\cite{Aebischer_2021,Kley:2021yhn}. Moreover, their effect is suppressed by the electron Yukawa coupling $y_e$, yielding to a negligible contribution proportional to $v^2 \,y_e/\Lambda_{\rm NP}^2$.
In Eq.~\ref{eq:eight}, 
the sine of the weak mixing angle is 
defined as $\sw^2=\frac{1}{2} \left( 1 - \sqrt{1 - 2 \sqrt{2}  \pi  \alpha /G_\mu M_Z^2} \right)$. The effective shifts $\Delta g^{Ze}_{L(R)}$ parametrize the dimension-six SMEFT deviations from the SM left- and right-handed couplings of the $Z$ boson to electrons, namely
$\hat{g}_L=\sw^2 -1/2$ and $\hat{g}_R=\sw^2$, see the first panel of Figure~\ref{fig:HNPdiags}. Their expression in the Warsaw basis~\cite{Grzadkowski:2010es}, given in the Appendix~\ref{APP:WCs}, is a linear combinations of WCs acting as effective $eeV$ couplings due to new interactions or modifications of the EW symmetry breaking entering the muon decay width when choosing $G_\mu$ as input
\footnote{We stress that if the electromagnetic coupling is not taken as an input also an overall 
$\Delta \alpha\neq 0$
shift is present at dimension six~\cite{Brivio:2020onw, Biekotter:2023xle}}.
At dimension six, also four-fermion contact interactions contribute to the Bhabha cross section, as shown in the second diagram in Fig.~\ref{fig:HNPdiags}. Their Lagrangian is independent of the input scheme and reads as follows 
\begin{equation}\label{EQ:4fermions}
    \begin{aligned}
\mathscr{L}_{\rm{SMEFT}}^{{4f}}=  \frac{1}{2}\frac{C_{ll}}{\Lambda^2_\text{NP}}&\left(\bar{e}_L \gamma^\mu e_L\right)\left(\bar{e}_L \gamma_\mu e_L\right) \\
   +\frac{C_{le}}{\Lambda^2_\text{NP}}& \left(\bar{e}_L \gamma^\mu e_L\right)\left(\bar{e}_R\gamma_\mu e_R\right)
   \\
   +\frac{1}{2}\frac{C_{ee}}{\Lambda^2_\text{NP}}&\left(\bar{e}_R \gamma^\mu e_R\right)\left(\bar{e}_R \gamma_\mu e_R\right)\, ,
       \end{aligned}
\end{equation}
where the $1/2$ factors take into account the exchange of two identical currents.

At LO SMEFT, the prediction for the Bhabha cross section is given by
\begin{equation}\label{EQ:sigmaSMEFT}
   \sigma_{\rm{SMEFT}} =  \sigma_\text{SM}+\sigma^{(6)}=\sigma_\text{SM} +  \sum_{i=1}^n\frac{C_i}{\Lambda_\text{NP}^2}\sigma_i^{(6)}
   \,,
\end{equation}
where $\sigma_i^{(6)}= 2 \Re \mathcal{M}^\dagger_\text{SM}\mathcal{M}^{(6)}_{{\rm{SMEFT}},i}$ is the interference between the SM and SMEFT amplitudes due to the $i$th operator, having factored out the dependency on the ratio of the WCs to the NP scale and assuming they are real. We set $\Lambda_{\rm{NP}}=\SI{1}{\tera\electronvolt}$, the EFT validity being ensured, as at small angle the relevant scale of the process is $\sqrt{|t|}=[s/2\,(1-\cos{\theta_{\rm{min}}})]^{1/2}$, which is at most $\order{{10^2}}~\rm{GeV}$ for $\sqrt{s}=\SI{3}{\tera\electronvolt}$ in the CLIC setup. We define the
central deviation from the SM (differential) cross sections due to heavy NP  and its $1\sigma$ uncertainty as
\begin{equation}(\delta\pm\Delta\delta)_{\rm{SMEFT}}=\frac{1}{\sigma_{\rm{SM}}}\hspace{-1mm}\left( \sigma^{(6)}\pm \sqrt{\sum_{ij}\sigma^{(6)}_iV_{ij}\sigma_j^{(6)}}\right)\hspace{-1mm},
\label{eq:deltasmeft}
\end{equation}
where the covariance matrix $V_{ij}=\Delta C_i\rho_{ij}\Delta C_j$ takes into account the correlation between fitted coefficients and their errors. 
We use the flavor-general bounds on coefficients and correlation matrix from the global fit of Ref.~\cite{Falkowski:2017pss}, that include EW and low-energy precision data. The values of WCs are reported in Table~\ref{tab:WC_fit}  
\begin{table}[h!]
\centering
\begin{tabular}{cc}
\hline \hline\addlinespace[2pt]
$C_i$                  & $C_i\pm \Delta(C_i)$           \\
\addlinespace[1pt]\toprule 
$\Delta g_L^{Ze}$& $-0.0038 \pm 0.0046 $ \\
$\Delta g_R^{Ze}$ & $-0.0054 \pm 0.0045$  \\
$C_{ll}$         & $0.17 \pm 0.06$       \\
$C_{le}$       & $-0.037\pm 0.036$      \\
$C_{ee}$       & $0.034\pm 0.062$     \\   
\hline \hline
\end{tabular}
\caption{\label{tab:WC_fit}Central values and $68\%$ uncertainty of the WCs, taken from the flavor general results of Eq.~(4.8) in Ref.~\cite{Falkowski:2017pss}.} 
\end{table}
whereas the correlation matrix is given in the Appendix~\ref{APP:WCs}. The shifts of the $Zee$ couplings are well constrained yielding a negligible contribution to the SABS cross section; therefore, we consider only the impact of four-fermion operators.
In Table~\ref{tab:SMEFT_numbers} we show the uncertainty due to heavy NP deviation for SABS in the luminometer setup of future colliders. It can be seen that possible BSM effects would affect the luminosity precision target of future $e^+ e^-$ facilities in a non-negligible way \footnote{We have verified that, by taking $\alpha(0)$ instead of $\alpha(M_Z)$ as input, the results change by a factor $\alpha(M_Z)/\alpha(0)$, which is well within the $68\%$ error $\Delta \delta_\text{SMEFT}$.}. 
With polarized beams at ILC and CLIC, we have checked that the results worsen up to a factor of 2, due to the enhancement of $C_{ll}$ over $C_{ee}$. 
We have also checked that NP uncertainties were well below the precision goal of the luminosity determination at LEP with SABS, also by considering the bounds on contact interactions given in~\cite{ALEPH:2013dgf}. At BELLE-II, which measures the machine luminosity using a large acceptance with Bhabha events, the effect is below the $10^{-3}$ precision budget. 
 \begin{table}
    \centering
    \begin{ruledtabular}
    \begin{tabular}{lcccr}
    Exp. & $[\theta_\text{min}, \theta_\text{max}]$  & $\sqrt{s}~[\SI{}{\giga\electronvolt}]$ &  $(\delta\pm \Delta\delta)_{\rm{SMEFT}}$&$\Delta L / L$\\\addlinespace[2pt]
    \toprule
\multirow{4.000000}{*}{FCC } & \multirow{4.000000}{*}{$[\ang{3.7},\ang{4.9}]$} &$91$ & $(-4.2 \pm 1.7 )\times 10^{-5}$ & $<10^{-4}$ \\
& &$160$  & $(-1.3 \pm 0.5 )\times 10^{-4}$&\multirow{3.000000}{*}{$10^{-4}$} \\
& &$240$ & $(-2.9 \pm 1.2 )\times 10^{-4}$&\\
& &$365$ & $(-6.7 \pm 2.7 )\times 10^{-4}$&\\
\midrule 
\multirow{2.000000}{*}{ILC}  & \multirow{2.000000}{*}{$[\ang{1.7},\ang{4.4}]$} &\multirow{1}{*}{250}& $(-1.2 \pm 0.5 )\times 10^{-4}$&\multirow{2.000000}{*}{$<10^{-3}$}   \\
 & &$500$ & $(-4.9 \pm 1.9 )\times 10^{-4}$ &\\
\midrule
\multirow{2.000000}{*}{CLIC }& \multirow{2}{*}{$[\ang{2.2},\ang{7.7}]$} &$1500$ & $(-9.7 \pm 3.9 )\times 10^{-3}$&\multirow{2.000000}{*}{$<10^{-2}$}\\
& &$3000$  & $(-4.2 \pm 1.7)\times 10^{-2}$&  \\
    \end{tabular}
\end{ruledtabular}
        \caption{Heavy NP contamination  to SABS as in Eq.~\eqref{eq:deltasmeft} at future $e^+ e^-$ facilities. ${\Delta L}/{L}$ represents the luminosity target precision.}
\label{tab:SMEFT_numbers}
\end{table}

\begin{figure*}
    \centering
    \includegraphics[width=0.8\linewidth]{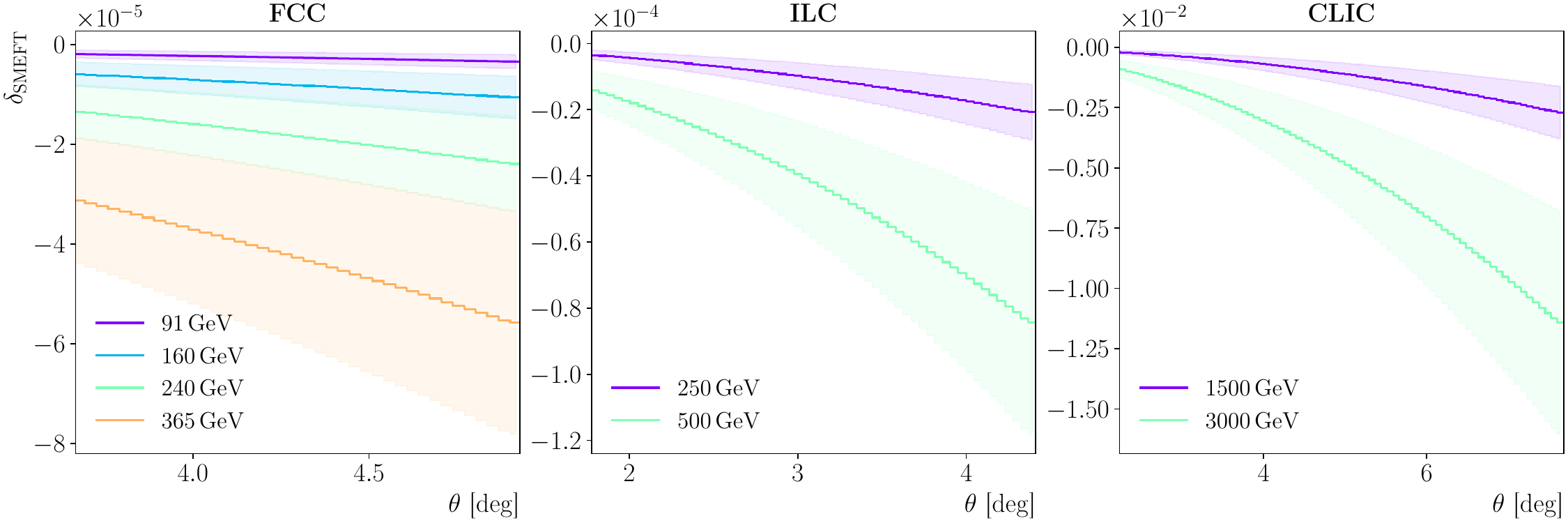}
    \caption{SMEFT deviation to the SABS differential cross section as a function of the electron scattering angle in the same setup of Table~\ref{tab:SMEFT_numbers}.}
    \label{fig:diff}
\end{figure*}

In Figure~\ref{fig:diff} we show the differential behavior of the SMEFT correction as a
function of the electron scattering angle, showing that the
deviations grow with $\theta_{e^-}$. This can be understood as the contact operators effect scales as $\delta_i^{4f}\simeq -sC_i(1-\cos\theta)/(\Lambda_\text{NP}^22\pi\alpha)$ with respect to the SM one. When reducing the angular acceptance, for example lowering the maximum angle to $\theta_\text{min}=\ang{4.2}$ in the $\sqrt{s}=\SI{91}{\giga\electronvolt}$ run of FCC-ee, the heavy NP contribution decreases to the $-2 \times 10^{-5}$ level, thus still representing a possible source of uncertainty.

At next-to-leading order (NLO) in the SMEFT the SABS prediction receives an additional contribution given by
\begin{equation}
  \sigma^{(6)}_\text{NLO}=\sum_j\frac{C_j(\mu)}{16\pi^2\Lambda_\text{NP}^2}2\Re\left(\Amp_\text{SM}^\dagger\Amp^{(6)}_{\text{NLO},j}(\mu)\right)
\end{equation}
where, in principle, the set of WCs entering the NLO prediction includes also operators not entering at LO accuracy. Systematic studies of NLO SMEFT corrections have been carried out for specific observables~\cite{Dawson:2019clf,Dawson:2021ofa,Asteriadis:2024xuk,Asteriadis:2024xts,Dawson:2024pft}, while global analysis has been performed at (partial) NLO accuracy in the SMEFT and including the effects of renormalization group evolution (RGE)~\cite{Bartocci:2024fmm,Celada:2024mcf}. 
We can estimate their effect on SABS by considering two different sets of contributions. 
%namely considering the NLO photonic correction to dimension-six effective operators or one loop SMEFT insertions on SM diagrams both in interference with the SM tree level. 
The former case, depicted in the third panel in Fig.~\ref{fig:HNPdiags}, is given by the NLO EW corrections to dimension-six operators and can be estimated as
\begin{equation}\label{nloov6}
    \frac{ \sigma^{(6)}_{\text{NLO},i}}{ \sigma^{(6)}_{\text{LO},i}}\simeq \frac{g_\text{SM}^2}{16\pi^2}\log\frac{M_Z^2}{|t|} \,.
    %\frac{\vert t \vert}{\Lambda_\text{NP}^2}\frac{C_i}{16\pi^2}\Delta_i\log\frac{M_Z^2}{|t|} 
\end{equation}
With $t$ the typical scale of the SABS in the various setup and $\Lambda_{\rm{NP}} \sim 1$~TeV, we conservatively get a contribution of the order of $\order{1 \%}$ 
relative to the LO SMEFT prediction.  
The latter case concerns NLO insertions of ${\hat O}_j$ operators not appearing at tree level, as shown in the last diagram in Fig.~\ref{fig:HNPdiags}. They contribute to the SABS cross section as 
\begin{equation}
     \frac{ \sigma^{(6)}_{\text{NLO},j}}{ \sigma_\text{SM}} \simeq   \frac{\vert t \vert}{\Lambda_\text{NP}^2} \frac{C_j}{16 \pi^2} \log \frac{\Lambda_\text{NP}^2}{\vert t \vert}\,,
\end{equation}
having chosen the renormalization scale as $\mu^2=\Lambda_\text{NP}^2$. This effect is expected to be more relevant at high energies, yielding to a contribution of $2\times 10^{-4}\,C_j$ with respect to the SM in the CLIC luminometer setup, while it is negligible in the other scenarios. Finally, RGE effects could give contributions of the same form of Eq~\eqref{nloov6}, with the replacement $\log M_Z^2/|t| \to \log M_Z^2/\Lambda_\text{NP}^2$, which are roughly of the same order of magnitude. 
For these reasons, we can safely estimate the bulk of the SMEFT contribution as given by operators appearing only at LO and neglecting their radiative corrections, which play a useful role in constraining directions in global fits but are not expected to alter our conclusions.

The future HL-LHC data are not foreseen to further constrain significantly the four-electron WCs, as discussed, for example, in~\cite{Celada:2024mcf}, in which the estimated reduction is of $\sim 20\%$ only. A possible strategy to remove the uncertainties induced by NP is to constrain the contact operators directly at future $e^+ e^-$ accelerators, by using observables that 
are independent of the absolute luminosity and can be 
measured with high precision.

\section{Constraining contact interactions with LABS}
In the worst-case scenario of no significant improvement on WCs bounds by the timeline of the start of future colliders, we explore in this section the possibility of constraining such coefficients. To this purpose, we rely on observables that do not depend on the luminosity and concern the LABS process, exploiting a phase-space region complementary to the SABS. Suitable quantities to this purpose are asymmetries, as they are defined as ratios and the luminosity dependence cancels out. They can be expressed in the following way:
\begin{equation}\label{eq:ASYdef}
      A_{ab}  =\frac{\sigma_a-\sigma_b}{\sigma_a+\sigma_b}\, ,
\end{equation}
where $\{a,b\}=\{\text{F},\text{B}\},\{\text{L},\text{R}\},\{\uparrow,\downarrow\}$ refer to different observables, as discussed in the following.
The theoretical prediction for $A_{ab}$ including the dimension-six SMEFT contribution can be written as
\begin{equation}\label{eq:ASYab}
    A_{ab}^\text{th}= A_{ab}^\text{SM}\left
\{1+\frac{(\sigma_{a}-\sigma_{b})^{(6)}}{(\sigma_{a}-\sigma_{{{b}}})_{\rm{SM}}}-\frac{(\sigma_{a}+\sigma_{b})^{(6)}}{(\sigma_{a}+\sigma_{b})_{\rm{SM}}}\right\}\, ,
\end{equation}
which depends on the WCs due to the definition of $ \sigma_{a,b}^{(6)}$ given in Eq.~\eqref{EQ:sigmaSMEFT}. This dependence can be exploited by fitting the above expression with MC simulated data assumed to be a Gaussian $ g\bigl(A^{{\rm{SM}}}_{ab},\Delta A_{ab}\bigr)_\alpha$ centered about the SM tree-level prediction. The error is computed as
\begin{equation}
\Delta A_{ab}=\sqrt{\sum_{k={a,b}}\left(\pdv{A_{ab}}{N_k}\right)^2\Delta_k^2}
=2\sqrt{\frac{N_a N_b}{(N_a+N_b)^3}}\, ,
\end{equation}
assuming only statistical uncertainty $\Delta_k=\sqrt{N_k}$ on the number of events, given by the product of the time integrated luminosity and the cross section for the process $N_a=L \sigma_a$.

Therefore, if asymmetries related to the LABS process can be exploited, no additional assumptions on the NP scenarios are needed, allowing one to fit Eq.~\eqref{eq:ASYab} with four-electron WCs. On the other hand, if one assumes lepton flavor universality also asymmetries related to $\mu^+ \mu^-$ and $\tau^+ \tau^-$ pair production could be used, providing more constraining power. Throughout this section we consider conservatively the LABS in the $\theta\in[\ang{40},\ang{140}]$ acceptance, in analogy with the event selection conditions of LEP analyses~\cite{Montagna:1998sp}. This allows us to enhance the $s$-channel contribution to the Bhabha cross section and the sensitivity to the SMEFT parameters, with the aim of constraining the four-electron $\vec{C}_{4f}=(C_{ll},C_{le},C_{ee})$ coefficients. We assume the validity of the SM up to the starting of future colliders physics programs and focus on the attainable uncertainties of the WCs by future asymmetry data~\footnote{Should the real data give values for the WCs not compatible with zero, this would be a clear indication of NP.}.

\subsection{The $Z$ resonance region}
\begin{figure}[h]
    \centering
\includegraphics[width=\linewidth]{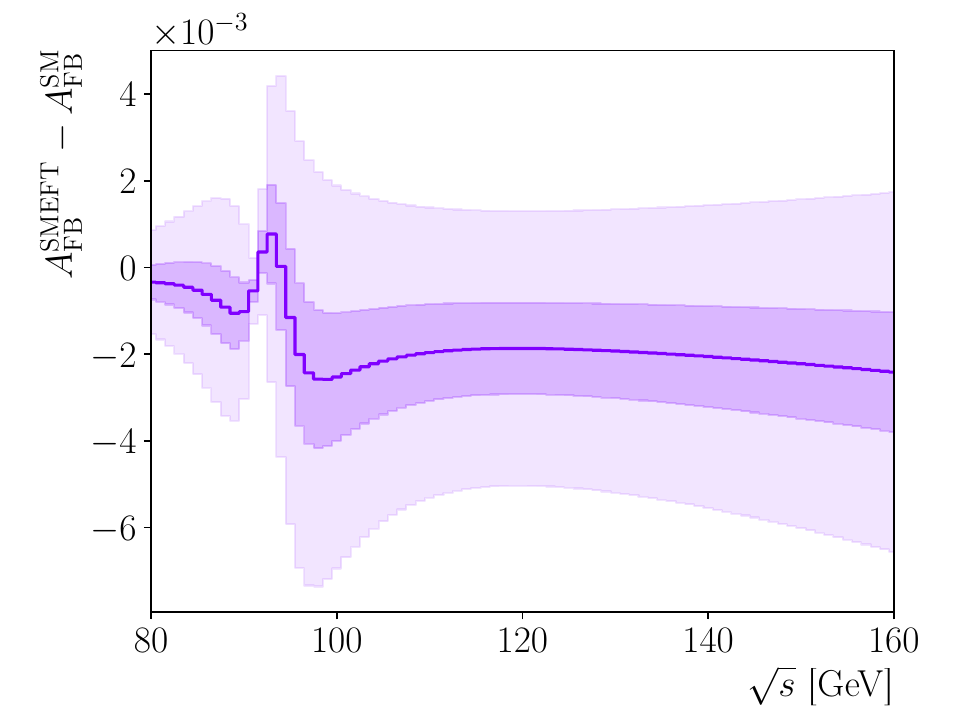}
    \caption{    \label{fig:NPdiff}Absolute deviation of the LABS $A_\text{FB}$ predicted in the SMEFT and the SM, using the current knowledge on WCs. The solid line is the central value whereas the shaded regions are the $1\sigma$ and $3\sigma$ uncertainty bands.}

\end{figure}

Around the $Z$ peak, the natural asymmetry to be considered is the forward-backward asymmetry $A_\text{FB}$, which in $s$-channel processes is a key quantity for the measurement of electroweak couplings and parameters~\cite{ALEPH:2005ab}. The $A_\text{FB}$ of the LABS scattering is defined as in Eq~\eqref{eq:ASYdef} where 
\begin{equation}
\sigma_{\rm{F}}=2\pi\int_0^c \dd \cos\theta \,  \frac{\dd\sigma}{\dd \Omega} \qquad \sigma_{\rm{B}}= 2\pi\int_{-c}^0 \dd \cos\theta \, \frac{\dd\sigma}{\dd\Omega}
\end{equation}
are the usual forwar and backward cross sections where the acceptance cut is set to $c=\cos\ang{40}=0.77$.
The measurement of $A_\text{FB}\left(\sqrt{s}\right)$ at three different values of $\sqrt{s}$ can provide sufficient information to simultaneously constrain the three independent $\vec{C}_{4f}$ coefficients. 
The three energy points more sensitive to deviations from the SM can be identified by calculating the difference between the prediction in the SM and in the SMEFT with current bounds on $\vec C_{4f}$, as shown in Fig.~\ref{fig:NPdiff} and they are found as the two local minima located at $\sqrt{s_1}=\SI{89}{\giga\electronvolt},\sqrt{s_3}=\SI{98}{\giga\electronvolt}$ and the maximum at $\sqrt{s_2}=\SI{93}{\giga\electronvolt}$. It is worth stressing that the radiative corrections will play an important role, in particular in the choice of the optimal energy points. Nevertheless we checked that, with the addition of QED leading logarithmic corrections switched on, the shape of the plot in Fig.~\ref{fig:NPdiff} is preserved, the largest shift affecting the second local minimum which moves to about $103$~GeV. By considering instead $\sqrt{s_3}=M_Z$ we still find enough constraining power on $\vec{C}_{4f}$.

The relative deviation of $A_{\rm{FB}}(\sqrt{s}_\alpha)$ in the SMEFT can be written by rearranging Eq~\eqref{eq:ASYab}~\cite{Berthier:2015gja}
\begin{equation}
 \label{eq:afb_system}
    \begin{aligned}
\sum_{i\in 4f} \frac{C_i}{\Lambda_\text{NP}^2}\left[\frac{(\sigma_{{\rm{F}}}-\sigma_{{\rm{B}}})_i^{(6)}}{(\sigma_{{\rm{F}}}-\sigma_{{\rm{B}}})_{\rm{SM}}}-\frac{(\sigma_{{\rm{F}}}+\sigma_{{\rm{B}}})_i^{(6)}}{(\sigma_{{\rm{F}}}+\sigma_{{\rm{B}}})_{\rm{SM}}}\right]_\alpha=\frac{\Delta A^{0}_{{\rm{FB}},\alpha}}{A^{0}_{{\rm{FB}},\alpha}}\, ,
\end{aligned}
\end{equation}
where the dependence on the WCs has been factored to the left while to the rhs we have the simulated data.
To quantitatively estimate the uncertainty reduction for four-electron WCs at FCC-ee or CEPC, we solve the system of Eq.~\eqref{eq:afb_system} by setting $\Delta A_\text{FB}$ to the foreseen experimental precision. Assuming six effective months of run ($5\times 10^6~ \mathrm{s}$)  for each energy point with a FCC-like instantaneous luminosity $\mathcal{L}_\text{FCC}=\SI{1.4e36}{\cm\tothe{-2}\second\tothe{-1}}$~\cite{FCC:2018byv}, we find that the statistical uncertainty on the forward-backward asymmetry is $\Delta A^0_{{\rm{FB}},\alpha}\lesssim \SI{5e-5}{}$. Therefore, the $1\sigma$ uncertainty on four-electron coefficients is reduced to $\Delta \vec{C}_{4f}\lesssim 10^{-2}$ yielding to a contribution $\delta_{\rm{SMEFT}} \simeq 5\times 10^{-6}$ on the $Z$-peak luminosity at FCC, which is smaller than the precision goal. This constraint would be enough also for future $e^+ e^-$ collider runs at higher energies. 
\subsection{The high-energy region}
\label{sec:apol}
In the scenario of high-energy future machines that will not revisit the $Z$ resonance scan, the $A_{\text{FB}}$ observable can not be used to constrain the four-fermion WCs, as it is almost constant for $\sqrt{s}\gtrsim\SI{120}{\giga\electronvolt}$. 
Since linear collider projects feature polarized beams, we investigate the possibility of using polarization asymmetries for LABS at $\sqrt{s}=\SI{250}{\giga\electronvolt}$~\footnote{{ Since we are considering a large-angle observable, going to higher energies could spoil the validity of the EFT. A low/intermediate energy run would therefore be important for precision measurements.}} at different angles assuming that they are independent of the luminosity~\cite{SLD:1994cex}. 
 For longitudinally polarized beams with fraction $P_{e^\pm}$ of polarized positrons and electrons, the differential cross section in the scattering angle can be written as
\begin{equation}\label{Eq:PolXsec}
   \dv{\sigma(P_{e^\pm}) }{\cos \theta}=\frac{1}{4} \sum_{\footnotesize{I,J={L,R}}}\left(1+ P_{e^+_{I}}\right)\hspace{-1mm}\left(1+P_{e^-_J}\right)\dd \sigma_{e^+_I e^-_J}
\end{equation}
where $P_{e^\pm_R}=P_{e^\pm}$ 
and $P_{e^\pm_L}=-P_{e^\pm}$, with the proposed values $|P_{e^-}|=0.8$ and $|P_{e^+}|=0.3$.
The theoretical prediction for polarization asymmetries $A^\text{th}_\text{pol}$, built from the SMEFT calculation of $\dd \sigma(P_{e^\pm})/\dd \cos\theta$ as defined in Eq.~\ref{Eq:PolXsec}, can be exploited to fit the confidence intervals of $\vec{C}_{4f}$. 
We consider $n=78$ experimental bins in $\cos\theta$ of width $\Delta\cos\theta=0.02$ and consider
a multivariate Gaussian likelihood~\cite{Berthier:2015gja} whose negative logarithm $\chi^2=-2\log L$ is defined as
\begin{equation}
    \chi^2 = -2\log L=\sum_{\alpha=1}^n\frac{\left(A_{\rm{pol}}^0-A_{\rm{pol}}^\text{th}(\vec{C}_{4f})\right)_\alpha^2}{(\Delta A_{\rm{pol}}^0)_\alpha^2}\, ,
\end{equation} 
which is distributed as a chi-squared with the number of d.o.f. given by $\nu=\text{dim}(\vec{C}_{4f})=3$.
The projected experimental value for the asymmetry in the $\beta$th bin and its statistical error $\Delta A_{\text{pol},\alpha}^0$ are computed as above, by taking an expected luminosity of $\mathcal{L}_{\rm{ILC}}=\SI{1.35e34}{\centi\metre\tothe{-2}\second\tothe{-1}}$ for six months of run ($\SI{5e6}{\second}$. 
$\chi^2(\vec C)=-2\log L(\vec C)$ Under these hypotheses, the $\chi^2$ can be rewritten as
\begin{equation*}
    \chi^2(\vec C) =\frac{1}{\Lambda_\text{NP}^4} \sum_{i,j} \sum_{\alpha,\beta} C_i\,  {\kappa}_{i,\alpha}^{(6)} \, W^{-1}_{\alpha\beta} \, {\kappa}_{j,\beta}^{(6)}\, C_j\, ,\end{equation*}
where $\kappa_{i,\alpha}^{(6)}=\partial A_{\text{pol},\alpha}^\text{th}/\partial C_i$ is the derivative of the SMEFT theoretical prediction. The matrix $V_{ij}^{-1}=\sum_{\alpha,\beta}  {\kappa}_{i,\alpha}^{(6)} \, W^{-1}_{\alpha\beta} \, {\kappa}_{j,\beta}^{(6)}$ is the inverse covariance matrix between WCs. The amplitude $\Delta(C_i)$ of the $68\%$ confidence interval on $C_i$ is given by $\sqrt{V_{ii}}$. This is equivalent to the projection on the $C_i$ axis of the region in the space of WCs corresponding to $\chi^2(\vec C)\leq 1$, which we obtain with MC replicas.

\begin{figure*}
    \centering
\includegraphics[width=0.8\linewidth]{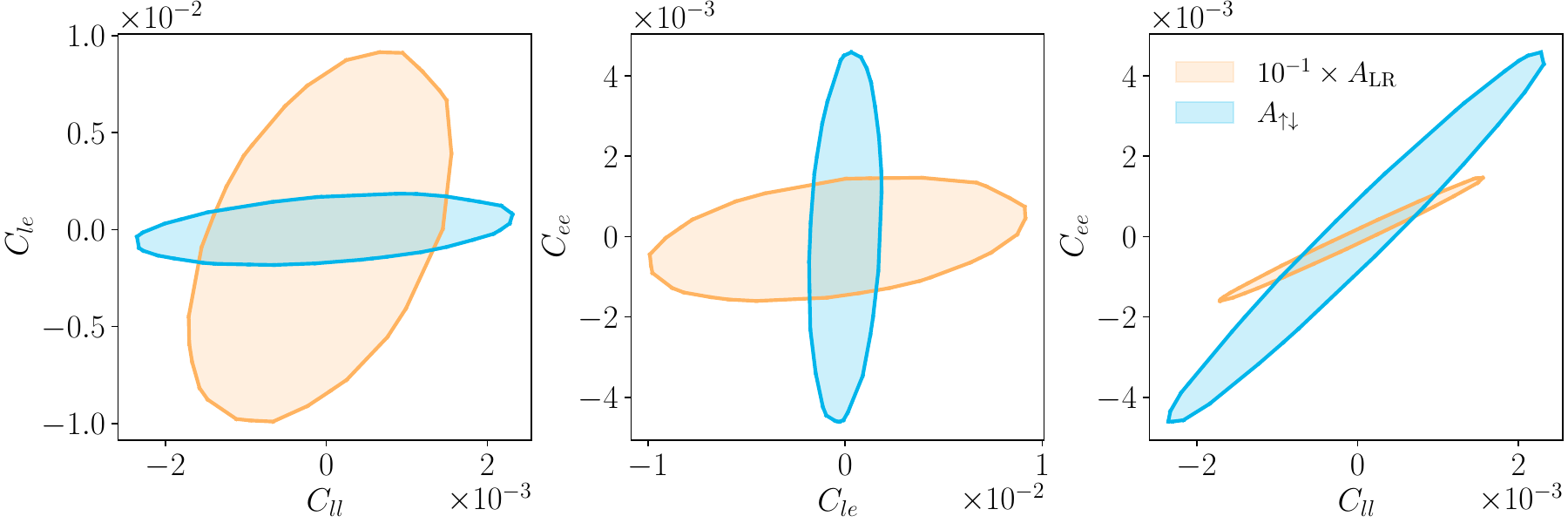}
    \caption{    \label{FIG:Ellipses}Results of the $\chi^2$ fit at $68\%$ confidence level projected in two-dimensional subspaces. As can be seen in the third panel, the fit of $A_{\rm{LR}}$ manifests an approximate flat direction in the $(C_{ll},C_{ee})$ plane. }
\end{figure*}

The commonly measured left-right asymmetry is defined as $A^{ff}_{\rm{LR}}=(\sigma_{\rm{L}}-\sigma_{\rm{R}})/(\sigma_{\rm{L}}+\sigma_{\rm{R}})$, where the right cross section $\sigma_{\rm{R}}$ is obtained by measuring the quantity of Eq.~\eqref{Eq:PolXsec} for a fixed set $(P_{e^+},P_{e^-})$ and the left one $\sigma_{\rm{L}}$ is obtained by inverting the sign of the polarizations. This observable was used at the SLD experiment at SLC for a precision measurement of $\sin^2\theta_{\rm{eff}}^\ell$~\cite{SLD:1997qsa,SLD:2000ujp} and it can be effective in future precision tests of the SM and of NP scenarios~\cite{ILC:2013jhg,ILCInternationalDevelopmentTeam:2022izu,Funatsu:2022spb,Miller:2024ivn}.
Unfortunately, as shown in the third panel in Fig.~\ref{FIG:Ellipses}, such asymmetry exhibits an almost exact flat direction as $C_{ll}\simeq C_{ee}$ due to its mild sensitivity to $C_{le}$, whose contribution
depends on the product $P_{e^+}P_{e^-}$ and, hence, cancels in the numerator of $A_{\rm LR}$. 
Therefore, in analogy with $A_\text{LR}$, we propose the following differential up-down asymmetry for a fixed $\sqrt{s}$ 
\begin{equation}
     \hspace{-1mm}  A_{\uparrow\downarrow}^-(P_{e^\pm},\cos\theta) = \frac{\dd\sigma(P_{e^+},P_{e^-})-\dd\sigma(P_{e^+},-P_{e^-})}{\dd\sigma(P_{e^+},P_{e^-})+\dd\sigma(P_{e^+},-P_{e^-})} \, ,
\end{equation}
inverting the sign of $P_{e^-}$. 
In Fig.~\ref{FIG:Ellipses} we show the results of the $\chi^2$ fit for $A_\text{pol}=A_{\rm{LR}},A_{\uparrow\downarrow}^-$ projected into 2D subspaces at $68\%$ confidence level. While $A_\text{LR}$ does not constrain further $\vec{C}_{4f}$, using the up-down asymmetry the uncertainty $\Delta C_i$ is reduced to a few $10^{-3}$ for all four-electron WCs. This would correspond to a negligible effect of $\delta_{\rm{SMEFT}}\lesssim 10^{-7}$ on the SABS luminosity measurements. 
\section{Conclusions}
\label{sec:conc}
Assessing the NP contamination to luminosity measurements will be crucial to achieve the precision physics program at future $e^+e^-$ colliders. In this paper, we have studied the BSM effects to the reference process SABS using an improved version of \textsc{BabaYaga@NLO}.

We have shown that the impact of light new scalars and vectors on the absolute luminosity is well below the 
expected experimental uncertainties for all scenarios.

The heavy NP contributions have been parametrized in a model-independent way, within the SMEFT framework. We have shown that the shifts of the SM $Zee$ couplings are irrelevant, whereas the uncertainties to the SABS due to four-electron operators are not negligible around the $Z$ resonance, as well as at higher energies.

In order to reduce these uncertainties below the luminosity target precision, we have proposed to make use of the LABS process and to rely on observables that are independent of the absolute luminosity. We have shown quantitatively
that the possible contamination around the $Z$ peak can
be removed by a measurement of $A_\text{FB}$ at three different
c.m. energy points close to the $Z$ resonance. These measurements would be enough also for the removal of any possible bias to higher energies.
In high-energy scenarios which do not foresee a run around the $Z$ resonance, the forward-backward asymmetry is not useful anymore.
However, high energy projects, like ILC and CLIC, feature beam polarization, which could allow to constrain
the WCs by means of polarization asymmetries. In particular, we have shown that the measurement of the left-right asymmetry is not sufficient to constrain simultaneously the three four-electron operators. However, an asymmetry that flips only on beam polarization, named $A_{\uparrow \downarrow}$, has the necessary constraining power.
The results of the present work could be put on firmer ground by including in our study the contribution of NLO corrections in the SM and SMEFT. This 
analysis is left to future work.

We also plan to further investigate the NP contamination to other relevant processes 
 for precision luminosity monitoring at future lepton facilities, such as two-photon production in $e^+e^-$ 
 annihilation and $\mu^+\mu^- \to \mu^+\mu^-$ at a muon 
 collider.

\section*{Acknowledgments}
We are indebted to Carlo M. Carloni Calame and Mauro Moretti for useful discussions. We thank Ilaria Brivio and Luca Mantani for helpful advice on SMEFT global fits and Marco Zaro for assistance with \textsc{MG5\_aMC@NLO}. The authors are grateful to Alain Blondel, Roberto Tenchini, Graham Wilson and Dirk Zerwas for fruitful feedback on the experimental side. We also acknowledge Juan Alcaraz Maestre for interest in our work. C.L.D.P. is supported by the U.S. Department of Energy under Grant No. DE-SC0012704. F.P. and F.P.U. thank the CERN theory department for hospitality during the completion of this work. 
\section{Data availability}
The data that support the findings of this article are not publicly available. The data are available from the authors
upon request.

\appendix
\section{Details on Wilson Coefficients}\label{APP:WCs}
In this appendix, we give further details on the WCs used in this work. We recall that the shifts to the $Zee$ SM couplings are defined in terms of the Warsaw basis as~\cite{Falkowski:2017pss}
\begin{eqnarray*}
\Delta g^{Ze}_{L} &=& \frac{1}{\Lambda_\text{NP}^2} \left\{ -\frac{1}{2} [C_{H l}]_{11}^{(3)} - \frac{1}{2} [C_{H l}]_{11}^{(1)} + f\left(-\frac{1}{2},-1\right) \right\} \\
\Delta g^{Ze}_{R} &=& \frac{1}{\Lambda_\text{NP}^2} \left\{ - \frac{1}{2} [C_{H l}]_{11}^{(1)} + f(0,-1) \right\} \, ,
\end{eqnarray*} 
with
\begin{equation*}
\begin{aligned}
    &f(T^3, Q) = - Q \frac{\sw \cw}{\cw^2 -\sw^2} C_{HWB} \\
+&\left( \frac{1}{4} [C_{ll}]_{1221} - \frac{1}{2} [C_{H l}^{(3)}]_{11} - \frac{1}{2} [C_{H l}^{(3)}]_{22} - \frac{1}{4} C_{HD} \right)\times\\ 
&\left( T^3 + Q \frac{\sw^2}{\cw^2 -\sw^2}\right) \, ,
\end{aligned}
\end{equation*}
where $T_3$ and $Q$ are the third component of the isospin and the charge of the electron, respectively, and the pedices indicate the flavor index $\{1,2,3\}=\{e,\mu,\tau\}$. The translation to other basis can be found in~\cite{LHCHiggsCrossSectionWorkingGroup:2016ypw}.

The correlation matrix between the shifts of the $Z$ couplings to electrons $\Delta g_{L/R}^{Ze}$ and the four-electron coefficients $\Vec{C}_{4f}$, whose numerical values are given in Table~\ref{tab:WC_fit}, is given by
\begin{equation*}
    \rho
    =\begin{pmatrix}
1 &  &  &  &  \\
0.15 & 1 &  &  &  \\
-0.09 & -0.08 & 1 &  &  \\
0.04 & -0.05 & -0.54 & 1 &  \\
0.08 & 0.08 & -0.04 & -0.54 & 1 \\
\end{pmatrix}\, .
 \end{equation*}

In this work we do not make any flavor hypothesis and use the results of~\cite{Falkowski:2017pss}. By adopting instead the results from the global fit in~\cite{Bartocci:2023nvp}, which assumes the more constraining $U(3)^5$ symmetry at high scale, the NP effect on the SABS cross section can be reduced approximately by a factor of 4, which is, however, not enough for the NP contamination to be below the precision goal at the future $e^+ e^-$ machines.

\bibliography{bibliography_inspire}

\end{document}